\documentclass[%
 reprint,
 amsmath,amssymb,
 aps,
]{revtex4-2}

\usepackage{graphicx}
\usepackage{dcolumn}
\usepackage{bm}
\usepackage{graphicx}
\usepackage{dcolumn}
\usepackage{bm}
\usepackage{siunitx, xcolor}
\usepackage{booktabs, array, mathptmx, float, tabularx, booktabs, lipsum, amsmath,multirow}
\usepackage[utf8]{inputenc}
\usepackage[T1]{fontenc}

\begin{document}

\preprint{APS/123-QED}

\title{Efficient Cross-layer Thermal Transport with Atypical Glassy-like Phenomena in Crystalline CsCu$_4$Se$_3$}



\author{Jincheng Yue,$^{1,\#}$ Yanhui Liu,$^{1, \#}$ Jiongzhi Zheng,$^{2,3}$}
\email{jiongzhi.zheng@dartmouth.edu}

\affiliation{$^{1}$Institute of High-Pressure Physics, School of Physical Science and Technology, Ningbo University, Ningbo 315211, China} 
\affiliation{$^{2}$Thayer School of Engineering, Dartmouth College, Hanover, New Hampshire, 03755, USA} 
\affiliation{$^{3}$Department of Mechanical and Aerospace Engineering, The Hong Kong University of Science and Technology, Clear Water Bay, Kowloon, 999077, Hong Kong}

\thanks{\# These authors contributes equally.}

\begin{abstract}
Understanding lattice dynamics and thermal transport in crystalline compounds with intrinsically low lattice thermal conductivity ($\kappa_L$) is crucial in condensed matter physics. In this work, we investigate the lattice thermal conductivity of crystalline CsCu$_4$Se$_3$ by coupling first-principles anharmonic lattice dynamics with a unified theory of thermal transport. We consider the effects of both cubic and quartic anharmonicity on phonon scattering rates and energy shifts, as well as the diagonal and off-diagonal terms of heat flux operators. Our results reveal that the vibrational properties of CsCu$_4$Se$_3$ are characterized by strong anharmonicity and wave-like phonon tunneling. In particular, the strong anharmonic scattering induced by Cu- and Cs-dominated phonon modes plays a non-negligible role in suppressing particle-like propagation. Moreover, the coherence-driven conductivity dominates the total thermal conductivity along the $z$-axis, leading to an anomalous, wide-temperature-range (100-700 K) glassy-like thermal transport. Importantly, the significant coherence contribution, resulting from the coupling of distinct vibrational eigenstates, facilitates effective thermal transport across layers, sharply contrasting with traditional layered materials. 
As a result, the non-monotonic temperature dependence of coherences' thermal conductivity results from the combined effects of anharmonic scattering rates and anharmonic phonon renormalization.
Our work not only reveals the significant contributions from the off-diagonal terms of heat flux operators in crystalline CsCu$_4$Se$_3$, but also explains the non-monotonic relationship between wave-like thermal conductivity and anharmonic scattering, providing insights into the microscopic mechanisms driving anomalous heat transport.

\end{abstract}

\date{\today}


\maketitle

\section{\label{sec:level1} INTRODUCTION}
\par In recent decades, the drive to enhance the efficiency of thermoelectric devices, along with the growing need for thermal insulation shields and coatings, has fueled extensive research into materials with ultra-low thermal conductivity($\kappa_L$)~\cite{zhao2014ultralow, russ2016organic, massetti2021unconventional}. Therefore, a fundamental understanding of lattice dynamics and thermal transport mechanisms in crystalline compounds with intrinsically ultra-low $\kappa_L$ is crucial for both scientific research and technological advancements. Generally, ultra-low $\kappa_L$ is fundamentally linked to strong lattice anharmonicity, which is often driven by unique electronic and atomic configurations~\cite{li2018crystal, asher2020anharmonic}. These configurations include, but are not limited to, lone electron pairs~\cite{jana2016origin, zhang2023dynamic}, resonant bonding structures~\cite{lee2014resonant}, and host-guest interactions~\cite{tadano2015impact}. Therefore, it is desirable to search for novel potential material with ultra-low $\kappa_L$ and uncover the intrinsic lattice dynamics from theoretical perspectives.

\par Copper-based chalcogenides exhibit several significant advantages, including their inherently low thermal conductivity, low toxicity, cost-effectiveness, and environmental sustainability~\cite{basit2023recent, ma2020alpha, Yue2024}. In particular, the copper-based chalcogenides exhibit diverse anharmonic phonon dynamics and atomic behaviors due to the unique characteristics of Cu atoms, such as the liquid-like behavior or loose bonding~\cite{wang2021structure, yue2024ultralow}. Notable examples include the superionic conductor Cu$_2$Se$_{1-x}$S$_x$~\cite{zhang2020cu}, strong bonding heterogeneity like Cu$_{12}$Sb$_4$S$_{13}$~\cite{xia2020microscopic}, and rattling model observed in $o$-CsCu$_5$S$_3$~\cite{yue2024ultralow}. Previously, Wu et al. successfully synthesized CsCu$_4$Se$_3$ using the boron-chalcogen method and reported an ultra-low $\kappa_L$ of approximately 0.2 Wm$^{-1}$K$^{-1}$ at around 650 K~\cite{ma2021mixed}. They attributed the microscopic mechanism underlying ultra-low $\kappa_L$ to the bond multiplicity induced by a hierarchical structure, low Debye temperature, and low transverse sound velocity. In addition, the rigid [(Cu$^+$)$_4$(Se$^{2-}$)$_2$]$^-$ double layer dominates the Se$^{2-}$ sublattice in Cu$_2$Se, providing enhanced stability. It acts as an ion-blocking barrier that limits atomic migration, keeping it below the decomposition threshold. This unique characteristic enhances the potential significance of crystalline CsCu$_4$Se$_3$ as a thermoelectric material. However, a comprehensive theoretical model that clarifies the underlying microscopic mechanisms of thermal transport remains lacking, primarily due to lattice instabilities and the limitations of the conventional phonon-gas model.

\par The outstanding challenge in theoretically understanding phonon-related properties arises from the limitations of the conventional harmonic approximation framework, which fails due to strong lattice anharmonicity~\cite{tadano2015impact, tadano2015self, pal2021microscopic}. In highly anharmonic layered materials, the thermal and quantum fluctuations of atoms are too substantial for quasiharmonic phonon theory to accurately capture the lattice dynamics~\cite{tadano2022first}. To overcome the limitations of the harmonic approximation, self-consistent phonon (SCP) theory within the quasiparticle (QP) approximation, has been employed to non-perturbatively treat anharmonic renormalization effects~\cite{tadano2018first}. Additionally, in highly anharmonic compounds, the reduction in propagative lattice thermal conductivity often leads to an increase in diffusive thermal conductivity~\cite{allen1999diffusons}. More specifically, the particle-like phonon transport model for describing thermal transport typically breaks down in highly anharmonic materials when phonon linewidths exceed the inter-branch spacings~\cite{peierls1929kinetischen}. Subsequently, the two-channel thermal transport model potentially bridges the gap, facilitating the evaluation of thermal conductivity in compounds between crystalline solids and amorphous glasses~\cite{mukhopadhyay2018two, luo2020vibrational}. In particular, the unified transport theory proposed by Simoncelli et al ~\cite{simoncelli2019unified, simoncelli2022wigner} offers a comprehensive approach that captures both particle-like and wave-like transport behaviors, effectively uniting the transport characteristics of crystalline solids and amorphous glasses.

\par In this work, we comprehensively investigate the lattice dynamics and thermal transport properties of crystalline CsCu$_4$Se$_3$ using a first-principles-based unified theory of thermal transport within an anharmonic lattice dynamics framework. More specifically, we incorporate both third- and fourth-order anharmonicity into phonon linewidths and energy shifts, and account for both off-diagonal and diagonal terms of heat flux operators in determining thermal conductivity. Given the strong anharmonicity in crystalline CsCu$_4$Se$_3$, we employed a self-consistent phonon theory incorporating bubble and loop diagrams to non-perturbatively capture the frequency shifts. Our findings suggest that the Cu-dominated flat branches and localized Cs-dominated modes induce strong anharmonic scatterings, significantly suppressing particle-like propagation. Notably, the coherence-driven wave-like conductivity dominates the total thermal conductivity along the $z$-axis, leading to an anomalous glassy-like thermal transport over a wide temperature range. Additionally, we find that heat transport across layers is effectively facilitated through wave-like tunneling, driven by the coupling of distinct vibrational eigenstates. Most importantly, the non-monotonic temperature dependence of the coherences' thermal conductivity arises from the combined effects of anharmonic scattering rates and anharmonic phonon renormalization. Our study not only elucidates the important role of coherences' thermal conductivity in across-layer thermal transport but also offers a detailed analysis of the non-monotonic temperature dependence of its glass-like thermal conductivity.

\section{\label{sec:level1}METHODOLOGY}
\subsection{Density-functional theory calculations}
\par Density functional theory (DFT) calculations were performed using the Projector Augmented-Wave (PAW) method~\cite{blochl1994projector}, as implemented in the Vienna Ab initio Simulation Package (VASP)~\cite{kresse1996efficient, kresse1996efficiency}. The valence states were assigned the following electron configurations: Cs(5s$^2$5p$^6$6s$^1$), Cu(3d$^{10}$4s$^1$) and Se(4s$^2$4p$^4$). The exchange-correlation function within the generalized gradient approximation (GGA)~\cite{perdew1996generalized} was implemented using the Perdew-Burke-Ernzerhof (PBE) version (See Fig. S1 in the Supplemental Material (SM)~\cite{Supplemental} for the functions test and discussion). Meanwhile, we use the optB86b-vdW~\cite{klimevs2011van} to describe the van der Waals (vdW) interactions accurately. A plane-wave basis set with a kinetic energy cutoff of 600 eV was employed to expand the Kohn-Sham orbitals. The ionic positions and unit cell geometry were thoroughly optimized using a plane-wave cutoff energy of 600 eV and a 12 × 12 × 10 Monkhorst-Pack $k$-point mesh, with a tight convergence criterion of 10$^{-4}$ eV·Å$^{-1}$ for forces and 10$^{-8}$ eV for total energy. The calculated relaxed lattice constants are $a$ = $b$ = 4.12 Å and $c$ = 10.25 Å, which shows good agreement with the experimental values ($a$ = $b$ = 4.09 Å and $c$ = 10.07 Å)~\cite{ma2021mixed}.

\subsection{Calculating interatomic force constants}
\par To calculate the interatomic force constants (IFCs), we use a supercell approach with a 
4 × 4 × 2 supercells, containing 256 atoms, for crystalline CsCu$_4$Se$_3$. The harmonic IFCs were calculated by performing a least-squares fit~\cite{tadano2015self} to displacement patterns obtained from the finite displacement method~\cite{esfarjani2008method, tadano2015impact}, using atomic displacements of 0.01 Å and 4 × 4 × 3 Monkhorst-Pack electronic $k$-point mesh.
Additionally, the potential energy surfaces were mapped using the same supercell dimensions and DFT calculation settings throughout. The anharmonic IFCs were subsequently calculated using the compressive sensing lattice dynamics (CSLD) method~\cite{zhou2014lattice}, where a compressive sensing matrix was generated from displacements combining trajectories of ab initio molecular dynamics (AIMD) simulations with additional random displacements. The AIMD simulation was conducted in the NVT ensemble at 300 K using a Nosé-Hoover thermostat, running for 6,000 steps with a 2 fs time step. To generate physically reliable atomic displacements, we initially sampled 100 atomic structures at regular intervals, excluding the first 1,000 steps from the AIMD trajectories. A random displacement of 0.1 Å was then applied uniformly to all atoms in random directions to minimize cross-correlations between structures. The training and cross-validation dataset for predicting the anharmonic IFCs was obtained from high-accuracy DFT calculations on the sampled atomic structures, with an energy convergence criterion better than 10$^{-8}$ eV. The compressive sensing method employs the adaptive least absolute shrinkage and selection operator (LASSO) technique ~\cite{nelson2013compressive} to optimize a sparse representation of higher-order anharmonic terms, incorporating contributions up to the sixth order in the Taylor expansion of the potential energy surface. Correspondingly, real-space cutoff radii of 8.46 Å, 7.41 Å, 4.23 Å, and 3.17 Å were set for the extraction of cubic, quartic, quintic, and sextic-order IFCs, respectively. In this work, all IFCs were determined using the ALAMODE package~\cite{tadano2015self}.

\subsection{Anharmonic phonon renormalization and unified thermal transport theory}
\par With the zero-K harmonic and anharmonic IFCs available, anharmonic phonon energy renormalization was carried out using self-consistent phonon (SCP) theory, as implemented in the ALAMODE package~\cite{tadano2015self}. The first-order self-consistent theory, which considers only quartic anharmonicity, can be expressed as~\cite{tadano2018first}
\begin{equation}
\begin{aligned}
(\Omega _q^S)^2=&(C^\dagger_q \Lambda _q^{(\mathrm{HA} )}C_q  )_{\nu \nu} \\& +\frac{1}{2N_{q'}} \sum_{q'} \Phi_4(q;-q;q';-q') \times \frac{\hbar }{2\Omega _{q'}^S}[1+2n(\Omega _{q'}^S)] 
\end{aligned}
\end{equation}
where $\Omega ^{S}_q$ is anharmonic phonon frequency with the wavevector $q$; $S$ represents the first-order SCP renormalization; $N_{q'}$ is the total number of
sampled phonon wavevectors in the first Brillouin zone; $\hbar$ is reduced Planck constant; $C_q$ is the unitary matrix used to obtain the corresponding polarization vectors, $\Lambda _q^{(\mathrm{HA} )}$ is the zero-K bare harmonic frequency matrix, defined as $\Lambda _q^{(\mathrm{HA} )}$ = diag($\omega _{q1}^2...\omega _{qs}^2$), $n(\omega)$ is the Bose-Einstein distribution function, $\Phi_4(q;-q;q';-q')$ is the representation of the fourth-order IFCs in the reciprocal space~\cite{tadano2015impact}. The second term on the right in Eq.(1) is the anharmonic self-energy associated with the loop diagram~\cite{tadano2015impact}, which can be considered as ${\textstyle \sum_{q}^{L}} [G,\Phi _4]$], where the $G$ stands for phonon Green’s functions~\cite{tadano2015self}.
Given the significant anharmonic polarization mixing (PM), we explicitly accounted for it by incorporating the off-diagonal components of the loop diagram self-energy. To obtain the renormalized phonon energies, $C_q$ and $\Lambda _q^{(\mathrm{HA} )}$ must be iteratively updated until the frequency meets the convergence criteria.

\par Considering the significant negative phonon energy shifts due to cubic anharmonicity in highly anharmonic compounds~\cite{tadano2015self, tadano2015impact}, we additionally incorporate a bubble self-energy correction into the renormalized phonons~\cite{zheng2022anharmonicity}. Correspondingly, based on the first-order renormalized phonon energies, the following self-consistent equation can be used to evaluate the further renormalized energy with the bubble diagram correction, labeled as the SCPB model, and is given by~\cite{tadano2022first}
\begin{equation}
\Omega ^2_{q}=(\Omega ^{S}_q)^2-2\Omega ^S_qRe\Sigma^B_q[G^S,\Phi _3](\omega )
\end{equation}
where ${\textstyle \sum_{q}^{B}} [G^S,\Phi _3](\omega)$ is the bubble self-energy on the basis of first-order SCP, which defined as
\begin{equation}
\begin{aligned}
 {\textstyle \sum_{q}^{(B)}} (\omega )=&\frac{1}{16}\sum_{q_1,q_2}\frac{\hbar |\Phi _3(-q;q_1;q_2)|^2}{\omega _q^S\omega _{q1}^S\omega _{q2}^S }\\& \times \Delta (-q+q_1+q_2)f(1,2,\omega +i\epsilon )  
\end{aligned}
\end{equation}
Here, the $\omega$-dependent function $f$ defined as~\cite{tadano2015self}
\begin{equation}
\begin{aligned}
f(1,2,\omega +i\epsilon )=&\sum_{\sigma =-1,1}\sigma [\frac{1+n(\omega^S_{q1})+n(\omega^S_{q2})}{\omega +i\epsilon+\sigma(\omega^S_{q1}+\omega^S_{q2})}\\&- \frac{n(\omega^S_{q1})-n(\omega^S_{q2})}{\omega +i\epsilon+\sigma(\omega^S_{q1}+\omega^S_{q2})}] 
\end{aligned}
\end{equation}
\par where $\Phi _3(-q;q_1;q_2)$ is the representation of the third-order IFCs in the reciprocal space~\cite{tadano2015impact}; $\Delta(q)$ is a function that becomes 1 when $q$ is an integral multiple of the reciprocal lattice vector and becomes 0 otherwise; $\omega +i\epsilon$ is the analytic continuation of Green's function to the real axis with a positive infinitesimal $\epsilon$~\cite{tadano2015self}. For the $q$ point in Eq. (1), we used the gamma-centered 4 × 4 × 2 grid. For the inner $q'$ point in Eq. (1) and the $q_1$ point in Eq. (3), we employed a denser 8 × 8 × 4 $q$-point mesh, with which we obtained convergence of anharmonic frequencies.
\par The conventional Peierls-Boltzmann theory~\cite{peierls1929kinetischen} considers only the diagonal terms of the harmonic heat-flux operator~\cite{srivastava1981diagonal}, neglecting off-diagonal contributions that become crucial in systems with complex crystal structures, strong anharmonicity, or disorder~\cite{chen2015twisting, pal2021microscopic}. Therefore, we adopted the Wigner transport equation (LWTE)~\cite{simoncelli2019unified, simoncelli2022wigner, caldarelli2022many}, which explicitly incorporates both diagonal and off-diagonal terms:
\begin{equation}
\begin{aligned}
\kappa _L=&\frac{1}{N_qV}\sum_{qjj'} \frac{c_{qj}\omega_{qj'}+c_{qj'}\omega_{qj}}{\omega_{qj}+\omega_{qj'}}\upsilon_{qjj'}\otimes\upsilon_{qj'j} \\& \times \frac{\Gamma _{qj }+\Gamma _{qj'}}{(\omega _{qj}-\omega _{qj'})^2+(\Gamma _{qj}+\Gamma _{qj'})^2} 
\end{aligned}
\end{equation}
\par where the $c_{qj}$, $V$, and $N_q$ are the mode heat capacity with the wavevector $q$ and the mode index $j$, primitive-cell volume, and the number of samples, respectively. The $\upsilon _{qjj'}=1/2(\omega _{qj}\omega _{qj'})^{-1/2}\left \langle \eta _{qj'}|\partial _qC(q)|\eta _{qj} \right \rangle$ represent the generalized interband group velocity with $C(q)$ and $\eta _{qj}$ being the dynamical matrix and polarization vector, respectively~\cite{tadano2022first}. The diagonal term ($j=j'$) corresponds to the Peierls contribution ($\kappa^P_{L}$) within the relaxation-time approximation, while the off-diagonal term gives the coherent contribution ($\kappa^C_{L}$); the total LTC is given as $\kappa_L$ = $\kappa^P_{L}$ + $\kappa^C_{L}$. The $\Gamma_{q}$ stands for the total linewidths including three-phonon (3ph), four-phonon (4ph), and isotope-phonon scattering processes. The phonon linewidths associated with 3ph scattering and 4ph scattering processes was computed as~\cite{tadano2015impact, tripathi1974self}
\begin{equation}
\begin{aligned}
\Gamma_q^{3ph}&=\mathrm{Im}{\textstyle \sum_{q}^{3ph}} [G,\Phi_3](\omega_q) \\&=\frac{\pi \hbar^2}{18}\sum_{q_1q_2q_3}\frac{\left | \Phi _3(-q;q_1;q_2) \right |^2 }{\omega _q\omega _{q1}\omega _{q2}} \Delta (-q+q_{1}+q_{2})  
\\&\times [ (n_1+n_2+1) \delta (\omega _q-\omega _{q1}-\omega _{q2})\\&-2(n_1-n_2)\delta (\omega _q-\omega _{q1}+\omega _{q2})]    
\end{aligned}
\end{equation}
\begin{equation}
\begin{aligned}
\Gamma_q^{4ph}&=\mathrm{Im}{\textstyle \sum_{q}^{4ph}} [G,\Phi_4](\omega_q) \\&=\frac{\pi \hbar^4}{96}\sum_{q_1q_2q_3}\frac{\left | \Phi _4(-q;q_1;q_2;q_3) \right |^2 }{\omega _q\omega _{q1}\omega _{q2}\omega _{q3}} \Delta (-q+q_{1}+q_{2}+q_{3})  
\\&\times [ (n_1n_2+n_2n_3+n_3n_1+n_1+n_2+n_3+1) \\&\delta (\omega _q-\omega _{q1}-\omega _{q2}-\omega _{q3})\\&+3(n_1n_2+n_1n_3+n_1-n_2n_3)\\&[\delta (\omega _q-\omega _{q1}+\omega _{q2}+\omega _{q3})-\delta (\omega _q+\omega _{q1}-\omega _{q2}-\omega _{q3})]]
\end{aligned}
\end{equation}
\par where we symbolically denote $n(\omega_{qi})$ as $n_i$; $\mathrm{Im}\sum$ is the imaginary part of the phonon self-energy; $\delta$ is the Dirac delta distributions, which is used to enforce conservation of energy.

\par The influence of isotope scattering can be accounted for using the mass perturbation approach~\cite{tamura1983isotope}, where the resulting phonon linewidth is given by the following expression
\begin{equation}
\begin{aligned}
\Gamma_{qj}^{iso}(\omega )&=\frac{\pi }{4N_q} \omega_{qj}^2
\\&\times \sum_{q_1,j_1}\delta (\omega -\omega _{q_1j_1})\sum_{\kappa }g_2(\kappa) \left | e^*(\kappa ;q_1j_1)\cdot e(\kappa ;qj) \right |^2   
\end{aligned}
\end{equation}
Here $e(\kappa;qj)$ is the phonon eigenvector with the wavevector $q$ and the mode index $j$; $g_2$ is a dimensionless factor given by $g_2(\kappa )=\sum_{i}f_i(\kappa )(1-m_i(\kappa )/M_{\kappa})^2 $, where $f_i$ is the fraction of the $i$th isotope of an element having mass, and $M_{\kappa}={\textstyle \sum_{i}} f_im_i(\kappa )$ is the average mass, respectively.
\begin{figure*}[t]
    \begin{center}
    \includegraphics[height=11cm]{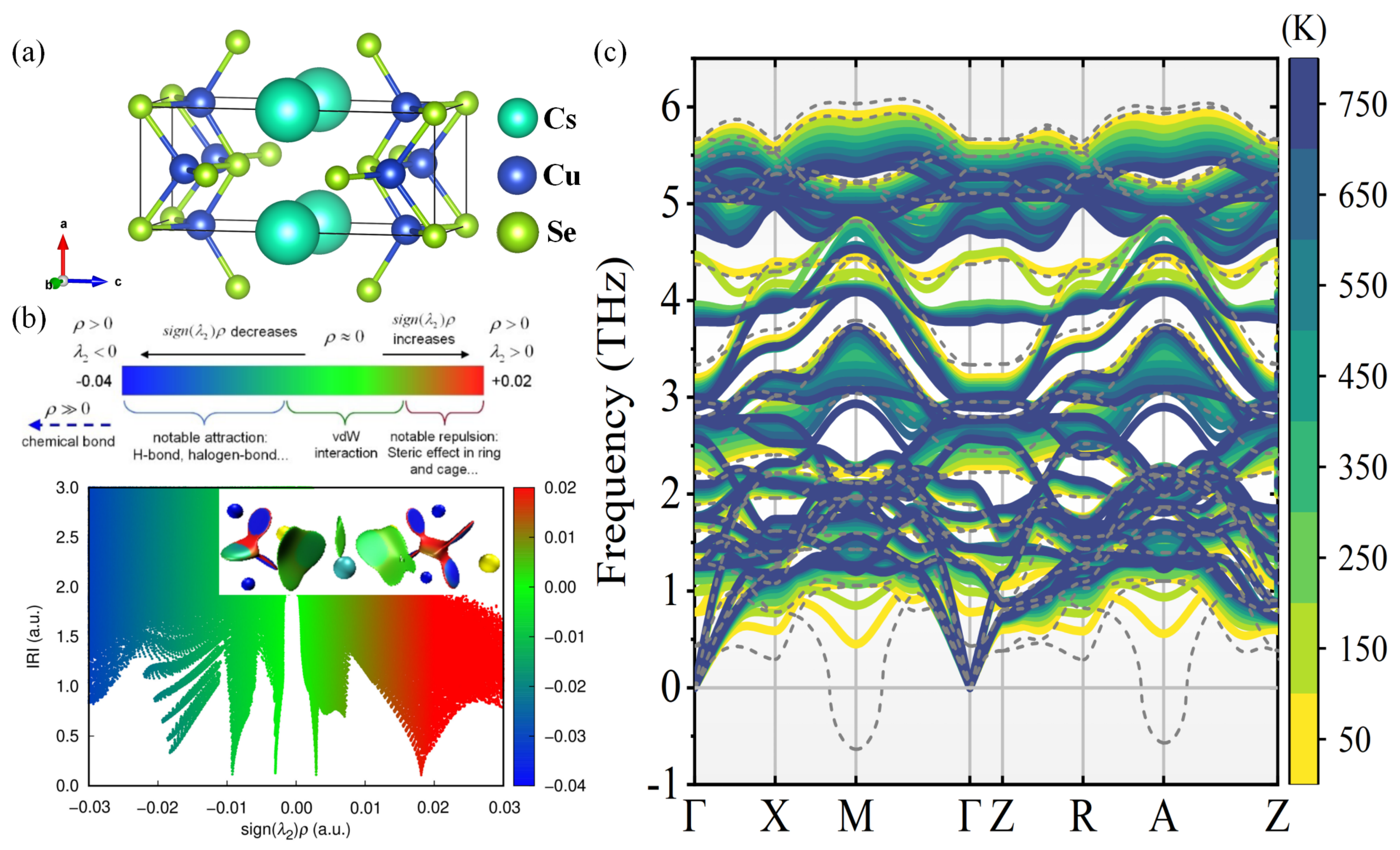}
    \caption{ (a) Crystal configuration of CsCu$_4$Se$_3$ with the unit cell marked visualized using VESTA. The cyan, blue, and green spheres represent cesium, copper, and sulfur atoms. (b) Non-covalent interaction analysis using the Interaction Region Indicator (IRI) method. Inset: Isosurface 2D map displayed with the standard coloring scheme and a chemical interpretation of the sign($\lambda_2$)$\rho$ on the IRI isosurfaces. The blue, green, and red regions characterize notable attraction, vdW interaction, and notable repulsion, respectively. (c) Harmonic (0K) and renormalized phonon dispersions using the self-consistent theory with bubble self-energy correction method at finite temperatures.}  
    \end{center}
 \end{figure*}
\par 
Calculating 4ph scattering probabilities and finding iterative solutions is challenging, requiring significant computation time and memory, and effective methods have been proposed in Ref.~\cite{ravichandran2018unified}. 
Nonetheless, incorporating 4ph scattering within the relaxation time approximation (RTA) is confirmed to yield negligible differences compared to the full solution of the Boltzmann transport equation (BTE)~\cite{ravichandran2020phonon}. In our calculations, considering the ultra-low $\kappa_L$ of CsCu$_4$Se$_3$, we applied the RTA methods, which have successfully been employed to predict the ultra-low thermal conductivity of various kinds of solids~\cite{carrete2014finding, seko2015prediction}.
In this work, the $q$-mesh for phonon scattering processes was set to 16 × 16 × 12 with a scale-broadening factor of 0.2 for the $\delta$ function, which gives well-converged results (The resulting error fluctuates around 1\%) for the crystalline CsCu$_4$Se$_3$ (See Fig. S2 in SM~\cite{Supplemental} for more discussion). Meanwhile, the maximum likelihood estimation proposed by Guo et al.~\cite{guo2024sampling} was used to accelerate the convergence of thermal conductivity. All thermal transport calculations incorporating particle-like propagation and wave-like tunneling transport channels were performed using the FourPhonon packages~\cite{li2014shengbte, han2022fourphonon}, as well as our in-house code~\cite{zheng2024unravelling}.

\begin{figure*}[t]
    \begin{center}
    \includegraphics[height=12.5cm]{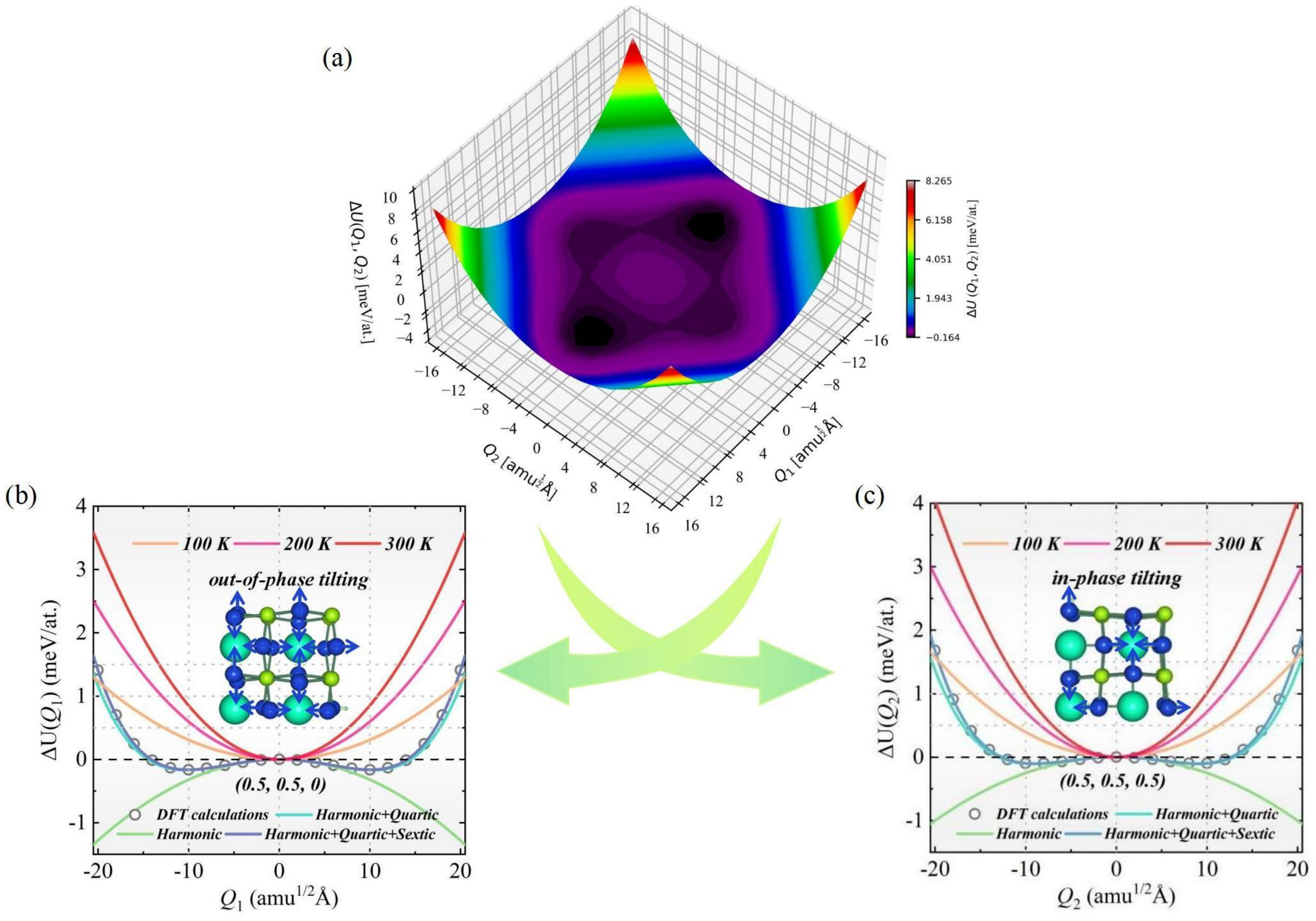}
    \caption{Calculated two-dimensional (2D) potential energy surfaces, and one-dimensional (1D) potential energy surfaces with corresponding displacement pattern. (a) Calculated 2D potential energy surface for CsCu$_4$Se$_3$ as a function of the normal mode coordinates $Q_1$ and $Q_2$. The soft modes at the $M$ and $A$ points, associated with their respective atomic eigenvectors, have been selected to generate the corresponding configurations. (b) DFT-calculated double-well 1D potential energy surface for the soft mode at the $M$ point, depicted as solid grey disks, as a function of the vibrational amplitude $Q_1$ in the normal mode coordinate. The potential energy surface is decomposed into the second (green dashed line), fourth (cyan dashed line), and sixth (blue dashed line) orders, respectively. 
    The orange, pink, and red curves are fitted using the renormalization frequencies via $2U(Q)=\omega_{qj} ^2Q_{qj}$ at 100, 200, and 300 K, respectively.
    (c) The same as (b), but for the $A$ point.} 
    \end{center}
 \end{figure*}

\section{\label{sec:level1}RESULTS AND DISCUSSION}
\subsection{Hierarchical configurations and bonding}
\par The crystal CsCu$_4$Se$_3$ adopts the $P4/mmm$ (No. 123) 2D hierarchical configuration consisting of the [(Cu$^+$)$_4$(Se$^{2-}$)$_2$]$^-$ double layers and Cs$^+$ ions slab, as shown in Fig. 1(a). Such a unique configuration is accompanied by an anisotropic behavior of the transport properties. It is worth mentioning that the Cu atom located at the 4$i$ site in CsCu$_4$Se$_3$ exhibits a two-fold rotational symmetry, in sharp contrast to the two-fold helical symmetry observed in Cu$_2$Se~\cite{ma2021mixed}. Therefore, the presence of Cs$^+$ cations effectively reduces the disorder of Cu atoms by decreasing the structural dimensionality, thereby enabling a higher aggregation density of the [(Cu$^+$)$_4$(Se$^{2-}$)$_2$]$^-$ double layers~\cite{ma2021mixed, yue2024ultralow}. Meanwhile, the hierarchical structure tends to induce weak bonding and bond multiplicity, which eventually gives rise to a large phonon anharmonicity~\cite{Yue2024, ma2019cscu5s3}.

\par To reveal the underlying bonding distribution, we calculated the Interaction Region Indicator (IRI)~\cite{lu2021interaction} as a function of $\mathrm{sign} (\lambda_2)\rho$ to quantify the atomic interaction. The IRI is defined by $\mathrm{IRI} (\mathbf{r} )=\left | \nabla \rho (\mathbf{r}) \right | /\left [ \rho (\mathbf{r}) \right ]^a $, where $a$ is an adjustable parameter (The standard definition of IRI is based on the adoption of a value of $a$ = 1.1~\cite{lu2021interaction}). The $\mathrm{sign} (\lambda_2)\rho$ represents the sign of the second-largest eigenvalue of the Hessian matrix of $\rho$~\cite{johnson2010revealing}, which can effectively distinguish between attractive and repulsive interactions~\cite{lu2012multiwfn}. Generally, the region with relatively high $\rho$, and consequently a large magnitude of $\mathrm{sign} (\lambda_2)\rho$, indicates a strong interaction (The blue area in Fig. 1(b)). Conversely, areas with low $\rho$, resulting in a small  $\mathrm{sign} (\lambda_2)\rho$, show the weak van der Waals (vdW) forces at most (The green area in Fig. 1(b)). As shown in Fig. 1(b), we observe multiple significant, shallow peaks near the zero thresholds within the low electron density regions ($\left | \mathrm{sign} (\lambda_2)\rho \right |  < $ 0.01 a.u.). This means that there are obvious weak vdW interactions in crystalline CsCu$_4$Se$_3$. To visualize the distribution of weak bonds, we plot the three-dimensional IRI isosurfaces (See the illustration of Fig. 1(b)). From the illustration of Fig. 1(b),  we notice that the weak van der Waals forces (green regions) are primarily concentrated around the Cs$^+$ ions. In contrast, the strong attractive interactions (blue regions) are symmetrically distributed between the Cu and Se atoms, highlighting the covalent bonding between them. As a result, such multiple bonding and weak bonding characteristics tend to drive strong lattice anharmonicity, as discussed below.

\begin{figure*}[t]
    \begin{center}
    \includegraphics[height=10cm]{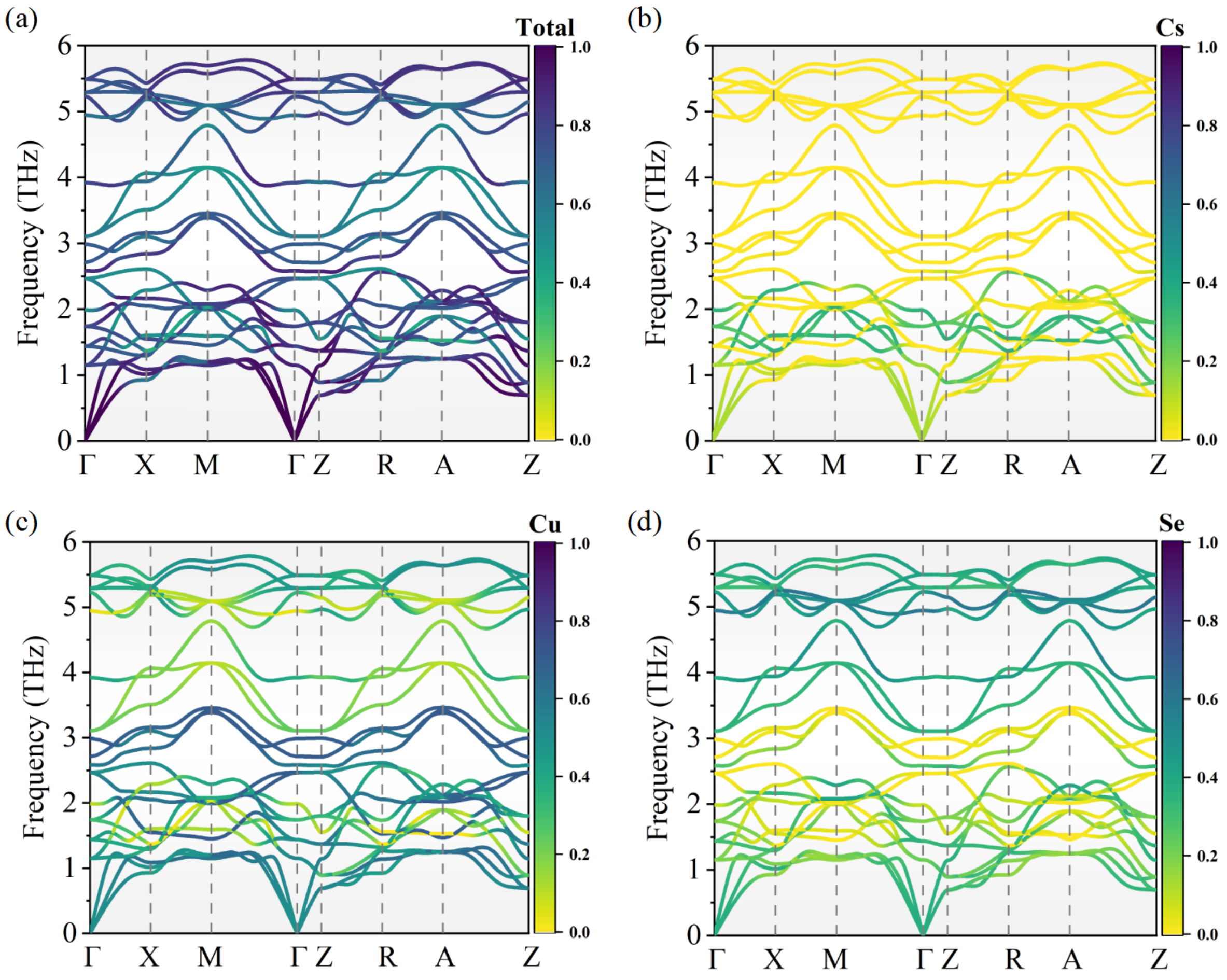}
    \caption{(a) Color-coded total atomic participation ratio (PR) projected onto the phonon dispersions at 300 K. (b) Color-coded atomic participation ratio (APR) of Cs atoms projected onto the phonon dispersions at 300 K. (c) The same as (b) but for Cu atoms. (d) The same as (b) but for Se atoms.} 
    \end{center}
 \end{figure*}

\subsection{Anharmonic self-energy and lattice dynamics}
\par Fig. 1(c) shows the phonon dispersions of CsCu$_4$Se$_3$ calculated using the harmonic approximation scheme and the self-consistent phonon calculation approach~\cite{tadano2015self, tadano2018first}, respectively. Clearly, when the anharmonic contributions to phonon dispersions are neglected, the zero-K harmonic phonon dispersions exhibit imaginary frequencies, particularly at the $M$ and $A$ points. Meanwhile, the imaginary phonon frequencies in crystalline CsCu$_4$Se$_3$ are primarily associated with the vibrations of Cu-atoms, as shown by the element-resolved phonon density of states (See Fig. S3(a) in SM~\cite{Supplemental}). To gain deeper insights into lattice instability and anharmonicity, we calculate the 2D potential energy surfaces (PES) as a function of the normal mode coordinates associated with the soft phonon modes (see Fig. 2(a)). Generally, the normal mode coordinates $Q_{qj}$ represent the amplitude of a particular vibrational mode. The mean square displacement of the normal coordinate $Q_{qj}$ is given as $<Q_{qj}^*Q_{qj}>=\frac{\hbar }{2\omega _{qj}} <A_{qj}^\dagger A_{qj}>=\frac{\hbar }{2\omega _{qj}}[1+2n(\omega _{qj})]$, where $A_{qj}$ stands for the displacement operator with crystal momentum $q$ and branch index $j$~\cite{tadano2015self}. 
More specifically, we create a series of crystal configurations with atomic displacements along normal modes at $q$-point in the specified supercell dimension, following the equation
\begin{equation}
\begin{aligned}
u_{qj} = \frac{Q_{qj}}{\sqrt{N_{\kappa}M_{\kappa}} }Re[exp(i\phi)e(\kappa;qj)exp(iq\cdot r_{\kappa l})] 
\end{aligned}
\end{equation}
where the $\phi$ is the phase; $N_{\kappa}$ is the number of atoms in the supercell and $M_{\kappa}$ is the atomic mass, $r_{\kappa l}$ is the position of the $\kappa$-th atom in the $l$-th unit cell. As expected, the minimum energy occurs away from the zero-tilt amplitude ($Q_1$ = $Q_2$ = 0) for both soft modes at the $M$ and $A$ points, as the indicator of the dynamical instability. The soft modes at the $M$ and $A$ points are associated with the out-of-phase and in-phase tilting of the Cu$_4$Se  units between adjacent layers, respectively, as shown in Figs. 2(b-c).

\begin{figure*}[t]
    \begin{center}
    \includegraphics[height=9.5cm]{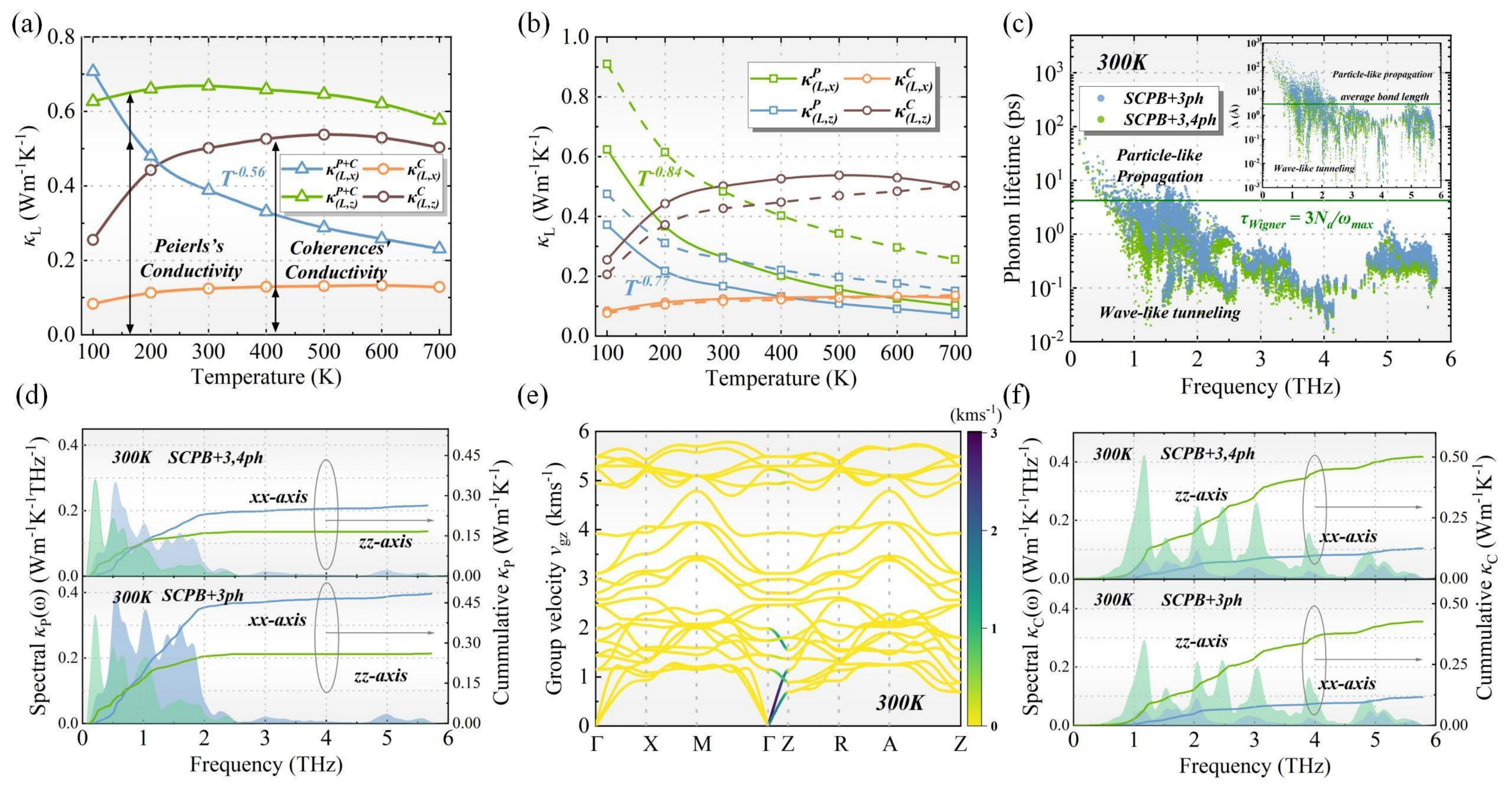}
    \caption{ (a) Calculated temperature-dependent total lattice thermal conductivity $\kappa_L$ and coherences' thermal conductivity $\kappa_L^C$ based on SCPB model, considering the three-phonon (3ph) and four-phonon (4ph) scattering along the $x$- and $z$-direction, respectively. 
    (b) Calculated temperature-dependent populations’ thermal conductivity $\kappa_L^C$ and coherences' thermal conductivity $\kappa_L^C$ based on SCPB model including the 3ph and 4ph scattering along the $x$- and $z$-direction, respectively. Dotted lines indicate that only 3ph scattering is considered, while solid lines indicate that both 3ph and 4ph scattering are considered.
    (c) Calculated phonon lifetime as a function of frequency at 300 K, where the green solid line represents the Wigner limit in time. ($\tau$ equals the inverse of the average interband spacing, i.e., $\tau_{Winger} = 3N_{at}/\omega_{max}$) Inset: Calculated phonon Mean free path (MFP) as a function of frequency at 300 K, where the green solid line represents the Ioffe-Regel limit in space. (MFPs equal to the typical interatomic spacing $a$.) 
    (d) Calculated spectral/cumulative populations’ thermal conductivity using the SCPB+3,4ph models including the 3ph and 4ph scattering along the $x$- and $z$-direction at 300 K, respectively. 
    (e) Projected phonon group velocity onto phonon dispersion along $z$-direction. (f) Calculated spectral/cumulative coherences’ thermal conductivity using the SCPB+3,4ph models including the 3ph and 4ph scattering along the $x$- and $z$-direction at 300 K, respectively. } 
    \end{center}
 \end{figure*}

\par Then we project the 2D PES onto 1D PES for the soft modes at the $M$ and $A$ points to gain a more comprehensive understanding of anharmonic lattice dynamics. Obviously, each soft mode exhibits the double-well potential due to the presence of lattice instability~\cite{zheng2024unravelling, zheng2022anharmonicity}. Moreover, the harmonic term fails to accurately represent the true shape of the PES, indicating the significance of the extremely strong lattice anharmonicity. Higher-order anharmonic contributions become more critical when collective atomic vibrations are substantial and non-negligible, especially under conditions of elevated high temperatures~\cite{tadano2018quartic, kim2023exploring}. Indeed, we can achieve good overall agreement with the DFT results by incorporating the contributions from the fourth-order IFCs. Meanwhile, including the sixth-order IFCs enhances the accuracy of the Taylor-expanded potential, even though the contributions from the sextic terms are relatively minor.
 
\par To accurately capture the intrinsic phonon scattering processes and the temperature dependence of phonon frequencies, it is crucial to incorporate the anharmonic terms in the lattice dynamics analysis. Accordingly, the renormalization process incorporates the third- and fourth-order anharmonic terms of the phonon self-energy, i.e., SCPB method, which includes the bubble (${\textstyle \sum_{q}^{B}} [G,\Phi _3](\omega)$) and loop (${\textstyle \sum_{q}^{L}} [G,\Phi _4]$) diagrams~\cite{tripathi1974self}. Only considering the ${\textstyle \sum_{q}^{L}} [G,\Phi _4]$ tends to overestimate phonon frequencies, whereas the negative frequency shift associated with ${\textstyle \sum_{q}^{B}} [G,\Phi _3](\omega)$ has been confirmed to be non-negligible~\cite{tadano2018quartic}. For CsCu$_4$Se$_3$, the high-order anharmonic terms cannot be treated as a perturbation of the non-interacting Hamiltonian $H_0$ due to the anharmonic terms being comparable with the harmonic term~\cite{tadano2018first}. Consequently, it is imperative to employ self-consistent methods that non-perturbatively address the anharmonic contributions to reflect the practical phonon frequency shift precisely. Indeed, the high-order anharmonicity was applied in the anharmonic phonon renormalization process to progressively convert the negative harmonic potentials of the soft modes into positive ones at finite temperatures (see Figs. 2(b-c)). These positive potentials reflect the effective harmonic potential, incorporating the influence of temperature.

\par After further consideration of the anharmonic renormalization, we discover that all soft modes significantly harden as the temperature increases. The soft modes at $M$ and $A$ points exhibit extremely sensitive temperature dependence, which illustrates the strong interaction with other phonons. On the contrary, the high-frequency phonon modes in the range 3-6 THz become softer with the increasing temperature, which is also observed in perovskite BaZrO$_3$~\cite{zheng2022anharmonicity} and Cs$_2$AgBiBr$_6$~\cite{zheng2024unravelling}. Besides, The higher-order self-energy contributions substantially impact the mean square displacement (MSD) of atoms (See Fig. S4 in SM~\cite{Supplemental}). Physically, this effect dynamically evolves the MSD by continuously modifying both the potential energy surface and the corresponding eigenvectors. In general, taking into the ${\textstyle \sum_{q}^{L}} [G,\Phi _4]$ tends to suppress the MSD, whereas the incorporation of ${\textstyle \sum_{q}^{B}} [G,\Phi _3](\omega)$ generally lead to a quantitative increase in MSD. 

\par Considering the distinct layered structure and complex atomic bonding in crystalline CsCu$_4$Se$_3$, we delve into the intrinsic phonon characteristics associated with the elemental vibrational dynamics by introducing the participation ratio (PR) and atomic participation ratio (APR)~\cite{pailhes2014localization}. 
The PR and APR are given as
\begin{equation} 
PR_{qj}=(\sum_{\kappa}^{N_{\kappa}} \frac{|e(\kappa;qj)|^2}{M_{\kappa}})^2/N_{\kappa}\sum_{\kappa=1}^{N_{\kappa}} \frac{|e(\kappa;qj)|^4}{M_{\kappa}^2}
\end{equation}
\begin{equation} 
APR_{qj,\kappa}=\frac{|e(\kappa;qj)|^2}{M_{\kappa}}/(N_{\kappa}\sum_{\kappa=1}^{N_{\kappa}} \frac{|e(\kappa;qj)|^4}{M_{\kappa}^2}) ^{1/2}
\end{equation}
where the $e(\kappa;qj)$ stands for the phonon eigenvector with the wavevector $q$ and the mode index $j$, $\kappa$ is the atomic index, $M_{\kappa}$ represents the atomic mass, and $N_{\kappa}$ signifies the total number of atoms. In general, a PR close to 1 indicates propagative phonon modes, with most atoms participating. In contrast, a lower PR reflects localized modes, where only a few atoms exhibit significant displacement while others possess small displacement or remain nearly stationary. Fig. 3(a) reveals that the low- and intermediate-frequency phonons, within the range of approximately 1.5 to 2.0 THz, exhibit highly localized behavior, as evidenced by their low participation ratio. Generally, highly localized phonon modes often lead to strong anharmonicity and play a critical role in suppressing thermal transport~\cite{tadano2015impact, zheng2022effects}. As shown in Fig. 3(b), the vibration of Cs$^+$ cations is mainly confined to a few phonon modes in the 1.5-2.0 THz range (See Fig. S3(b) in SM~\cite{Supplemental}). More importantly, we observe that the contribution from other atoms is almost negligible within these modes, which aligns with the low total PRs observed in Fig. 3(a), indicating the independent random oscillations of the Cs$^+$ cations. As a consequence, the Cs-dominated low- and intermediate-frequency phonons with low participation ratios will strongly interact with heat-carrying phonons, acting as significant roadblocks to thermal transport. Additionally, Cs$^+$ cations are located between adjacent [(Cu$^+$)$_4$(Se$^{2-}$)$_2$]$^-$ layers, making their vibrations critical to heat transport along the $z$-direction (see Fig. 1(a)). However, the vibrations of Cs atoms in crystalline CsCu$_4$Se$_3$ are highly localized and confined to a few specific phonon bands, leading to a 'quiescent' state of Cs$^+$ cations in other phonon modes~\cite{zeng2022extreme}. As a result, thermal energy carried by particle-like phonons is primarily restricted within the [(Cu$^+$)$_4$(Se$^{2-}$)$_2$]$^-$ layers, favoring transport along the layer direction ($x$-direction).

\par In Fig. 3(c), the Cu atoms primarily dominate the acoustic phonon branches, as well as the low-frequency ($\sim$1.2 THz) and mid-frequency (2-3.5 THz) optical phonon modes. As a result, the majority of heat-carrying phonons are linked to the vibrational modes of Cu atoms. Interestingly, the Cu-dominated low-frequency phonon branches ($\sim$1.2 THz) exhibit an extremely flat behavior, as shown in Fig. 3(d). Such unique characteristics can enhance phonon scattering processes by easily satisfying both energy and momentum conservation simultaneously~\cite{li2015ultralow}. Therefore, the Cu atoms exhibit behavior analogous to the "rattling" of guest atoms in filled skutterudite YbFe$_4$Sb$_{12}$~\cite{zheng2022effects} and clathrate Ba$_8$Ga$_{16}$Ge$_{30}$~\cite{tadano2015impact}, occurring around ~1.2 THz. Notably, we observe several pronounced avoided-crossing points between the longitudinal acoustic (LA) branch and the low-energy optical (LLO) branch (See Fig. S5 in SM~\cite{Supplemental}). These points represent strong hybridization between the optical and acoustic phonon polarizations near the region~\cite{yue2024role}. In general, the presence of avoided-crossing points is anticipated to suppress the group velocity, which has been confirmed by previous inelastic neutron scattering experiments. Also, the hybridization behavior leads to the admixture of optical phonon eigenvectors into acoustic phonon modes, giving rise to enhanced scattering rates~\cite{li2016influence}. In contrast, the Se atoms mainly exhibit cooperative vibration with Cu atoms in both the acoustic and high-frequency optical phonon branches (3 to 6 THz) (see Fig. 3(d)). 
\begin{figure*}[t]
    \begin{center}
    \includegraphics[height=10cm]{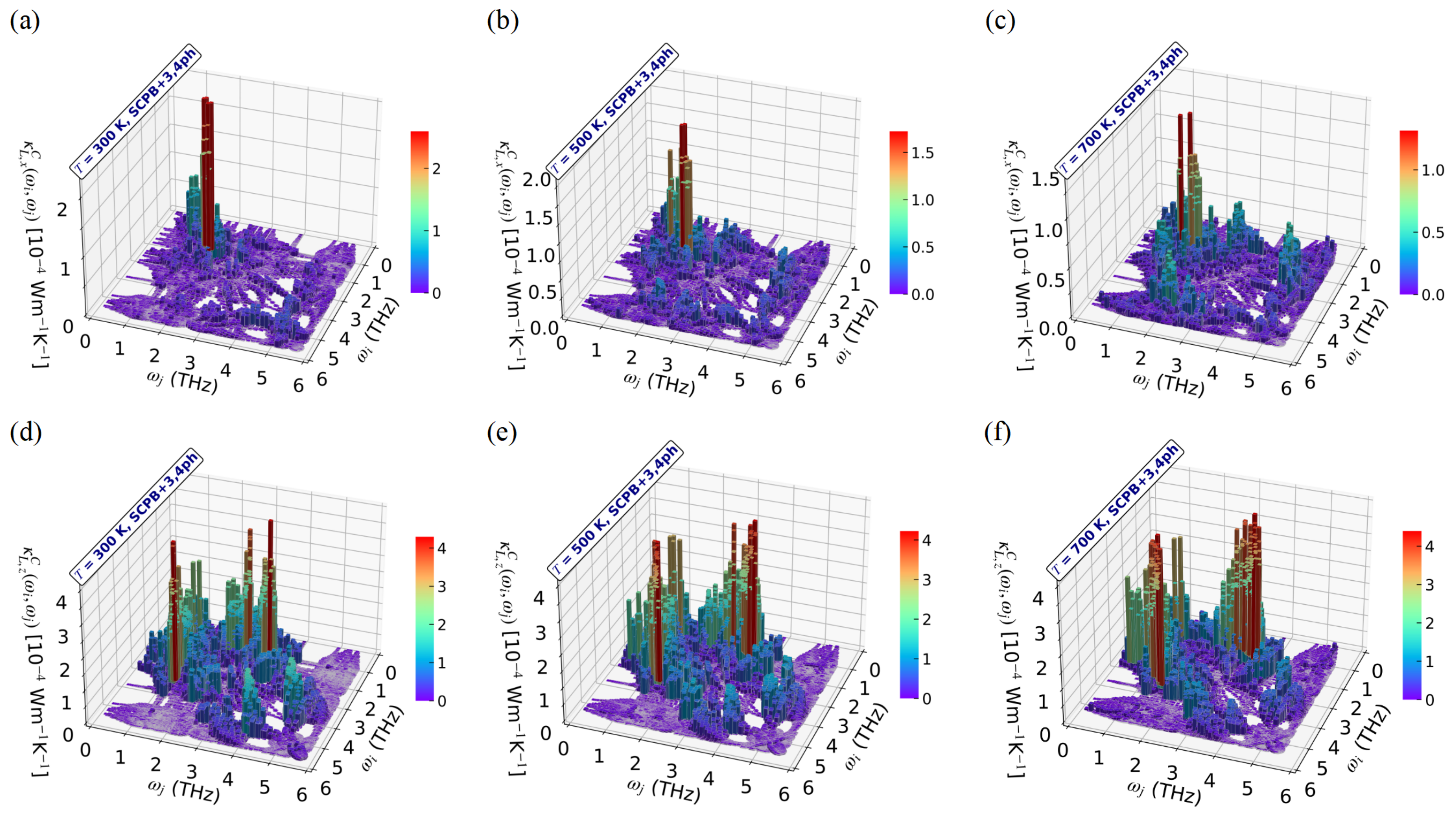}
    \caption{(a) Three-dimensional visualizations of the coherences' thermal conductivity $\kappa^C_L$($\omega_{qj}$,$\omega_{qj'}$) based on the SCPB+3,4ph model along with the $x$-axis at 300 K. The diagonal data points ($\omega_{qj}$ = $\omega_{qj'}$) indicate phonon degenerate eigenstates. (b) The same as (a), but for 500 K. (c) The same as (a), but for 700 K. (d) The same as (a), but for $z$-direction. (e) The same as (d), but for 500 K.
    (f) The same as (d), but for 700 K.} 
    \end{center}
 \end{figure*}
\subsection{Lattice thermal conductivity}
Next, we investigate the lattice thermal conductivity ($\kappa_L$), incorporating both Peierls and coherence channels, while considering the effects of cubic and quartic anharmonic terms. The Peierls channel involves the diagonal elements ($j = j'$) of the velocity operator, capturing particle-like transport via phonon wave packets, also known as populations' thermal conductivity. The coherence channel considers the off-diagonal elements ($j \ne j'$) of the velocity operator, reflecting the wave-like nature and originating from the loss of coherence between pairs of vibrational eigenstates $j$ and $j'$, known as coherences' thermal conductivity~\cite{simoncelli2019unified, simoncelli2022wigner}. Using the dual-channel thermal transport model, the converged total $\kappa_L$ values are 0.39 Wm$^{-1}$K$^{-1}$ and 0.69 Wm$^{-1}$K$^{-1}$ at 300 K along the $x$- and $z$-axes, respectively, as shown in Fig. 4(a). The calculated average thermal conductivity ($\kappa_L^{\mathrm{av} }=1/3( {\textstyle \sum_{\alpha=1}^{3}}\kappa_{L,\alpha} )$) is 0.49 Wm$^{-1}$K$^{-1}$, which aligns perfectly with the experimental values (0.48 Wm$^{-1}$K$^{-1}$ at room temperature)~\cite{ma2021mixed}. Conversely, the $\kappa_P^{ave}$ at 300 K is only 0.3 Wm$^{-1}$K$^{-1}$, significantly lower than the experimental value, highlighting the limitations of the conventional BTE approach for highly anharmonic CsCu$_4$Se$_3$. Although our SCPB+3,4ph model successfully reproduces the experimental lattice thermal conductivity in the 300–500 K temperature range, minor discrepancies between the experimental and theoretical thermal conductivity are observed in the 500–700 K range (see detailed comparison and discussion in Fig. S6 in the SM~\cite{Supplemental}). The close agreement between experimental and theoretical lattice thermal conductivity highlights the importance of accurately incorporating lattice anharmonicity and coherent thermal transport to predict 
$\kappa_L$ in materials exhibiting strong anharmonic behavior. Meanwhile, the ultra-low lattice thermal conductivity in crystalline CsCu$_4$Se$_3$ is competitive with those of state-of-the-art typical thermoelectric materials, indicating its promising potential for applications in thermoelectrics. 

\par In sharp contrast to other layered materials, such as SnSe~\cite{zhao2016ultrahigh} and BiCuSeO~\cite{zhao2014bicuseo}, the out-of-plane lattice thermal conductivity ($\kappa_{L,z}$) is higher than the in-plane conductivity ($\kappa_{L,x}$) over a wide temperature range (see Fig. 4(a)). This anomalous observation can be attributed to the substantial coherence contribution along the $z$-axis, resulting from the complex structure and strong anharmonicity. More specifically, the coherence contribution $\kappa^C_{L,x}$ along the $x$-direction reaches 0.12 Wm$^{-1}$K$^{-1}$ at 300 K, but an impressive 0.5 Wm$^{-1}$K$^{-1}$ along the $z$-direction. Owing to the obvious disparity in coherence contributions along different axes in crystalline CsCu$_4$Se$_3$, the thermal conductivity exhibits distinctly different temperature dependencies along the two directions. The temperature dependence of $\kappa_{L,x}$ is notably weak, following a $T^{-0.56}$ trend, due to the combined influence of high-order anharmonic self-energies and wave-like tunneling channels. In comparison, $\kappa_{L,z}$ displays an unusual behavior of initially increasing and then decreasing, driven by the dominant contribution of the wave-like tunneling channel along the $z$-axis, as discussed later.

\par We next proceed to elucidate the microscopic mechanisms underlying thermal transport in crystalline CsCu$_4$Se$_3$. Fig. 4(b) shows that quartic anharmonicity significantly influences both populations' and coherence thermal conductivity. With the incorporation of 4ph scattering, $\kappa^P_{L,x}$ and $\kappa^P_{L,z}$ decreased from 0.48 Wm$^{-1}$K$^{-1}$ and 0.26 Wm$^{-1}$K$^{-1}$ to 0.26 Wm$^{-1}$K$^{-1}$ and 0.16 Wm$^{-1}$K$^{-1}$ at 300 K, respectively. As illustrated in Figs. 4(c-d), the reduction in particle-like thermal conductivity due to 4ph scattering rates can be attributed to shortened phonon lifetimes. Specifically, Fig. 4(c) demonstrates that including 4ph scattering significantly reduces the lifetimes of particle-like phonons, particularly for those with frequencies below 1.5 THz. Correspondingly, within the frequency range of 0–1.5 THz, the particle-like phonon conductivity decreases by 43\% due to the impact of four-phonon scattering, as shown in Fig.4(d). As previously discussed (see Fig. 3(c)), phonons with frequencies below 1.5 THz are primarily dominated by Cu atoms, forming flattened bands that contribute to strong 4ph scattering~\cite{li2015ultralow, tadano2015impact}. Furthermore, for both 3ph and 4ph scattering rates in Fig. 4(c), we observe three extended tails within the frequency range of approximately 1.0–1.5 THz, corresponding to two dips in the spectral thermal conductivity shown in Fig. 4(d). To intuitively demonstrate the influence of low-frequency Cu-dominated modes on $\kappa^P_{L}$, we investigate the variation in $\kappa^P_{L}$ by excluding the contribution of branches within the 1.0-1.5 THz in the 3ph scattering calculations (See Fig. S7(a) in SM~\cite{Supplemental}). The results exhibit that neglecting the contribution of Cu-dominated optical modes results in a substantial increase in $\kappa^P_{L,x}$, reaching approximately 100.3 Wm$^{-1}$K$^{-1}$ from 0.48 Wm$^{-1}$K$^{-1}$, at 300 K using the SCPB+3ph model (See Fig. S7(b) in SM~\cite{Supplemental}). This observation indicates that phonons contributed by Cu atoms exhibit strong anharmonic scattering rates, highlighting its crucial role in suppressing particle-like thermal conductivity.

\par On the other hand, we examine the impact of Cs-dominated phonons on particle-like phonon conductivity given their localized nature in the 1.5–2 THz range. As shown in Fig. 4(c), several deep dips in phonon lifetimes are evident, reflecting the characteristics of localized modes~\cite{zheng2022effects} and further suppressing thermal transport. Although the 4ph scattering contributed by Cs-dominated modes is not as significant as 3ph scattering, these Cs-dominated phonon modes still undergo notable 4ph scattering processes (see Fig. 4(c)). Specifically, in the 1.5–2 THz frequency range, including 4ph scattering leads to a 51\% reduction in particle-like phonon conductivity, as shown in Fig. 4(d). Similarly, we also investigate the variation in $\kappa^P_{L}$ by excluding the contribution of branches within the 1.5-2.0 THz in the 3ph scattering calculations to investigate the influence of Cs-dominated modes (See Fig. S7(a) in SM~\cite{Supplemental}). Interestingly, we found that neglecting Cs-dominated modes (1.5-2.0 THz) also leads to a significant increase in $\kappa^P_{L,x}$ to 83.2 Wm$^{-1}$K$^{-1}$ at 300 K (See Fig. S7(b) in SM~\cite{Supplemental}).
We, therefore, conclude that the Cs-dominated vibration modes are almost as important as the Cu-dominated mode in suppressing thermal conductivity in CsCu$_4$Se$_3$.

\par Despite emphasizing the strong anharmonic scatterings of Cs-dominant modes, the Cs$^+$ anions still play a decisive role in particle-like cross-layer thermal transport along the $\Gamma$-Z direction. As shown in Fig. 4(e), all modes with nonzero $\upsilon_z$ are concentrated exclusively along the $\Gamma$-Z direction. Notably, branches with larger $\upsilon_z$ are linked to collective vibrations with a higher APR of Cs$^+$ ions (see Fig. 3(b)). In contrast, $\upsilon_z$ is nearly 0 in the high-frequency branches that lack contributions from Cs$^+$ ions, despite the Cu and S atoms exhibiting strong vibrations. Compared to $\upsilon_z$, $\upsilon_x$ and $\upsilon_y$ exhibit high values, even in the absence of Cs$^+$ ions during the vibration (See Fig. S8 in SM~\cite{Supplemental}). Consequently, the $\upsilon_z$ is almost constrained to vibrational modes associated with Cs$^+$ ions along the $\Gamma$-Z direction, which demonstrates the importance of the Cs$^+$ ions in facilitating particle-like heat transport between layers (see Fig. 3(b)).

 \begin{figure}[!htpb]
    \begin{center}
    \includegraphics[height=7.5cm]{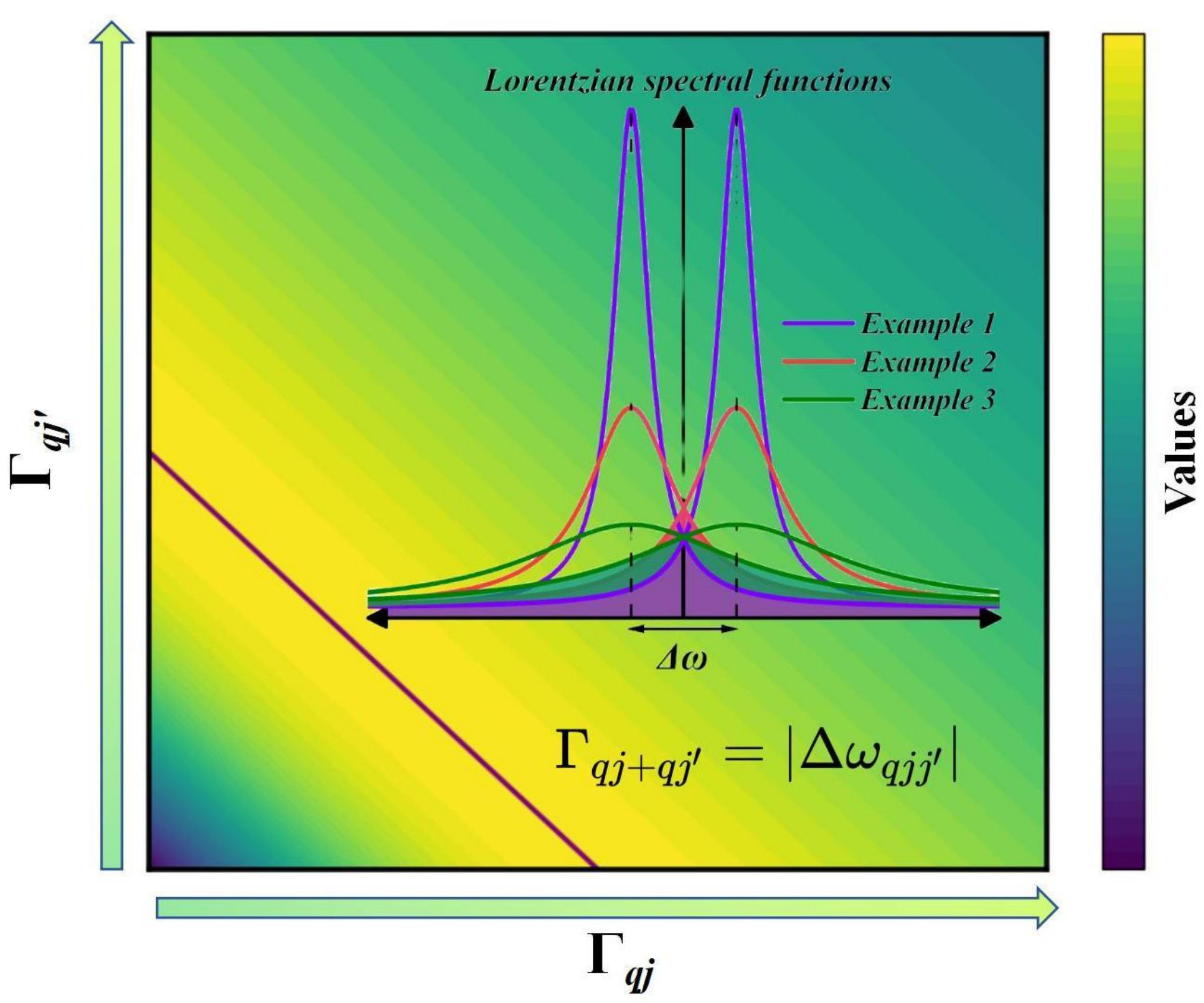}
    \caption{ The heat map of the $\Gamma _{qj+qj'}/(|\Delta\omega_{qjj'}|^2+(\Gamma _{qj+qj'})^2)$ as a function of total linewidth $\Gamma _{qj+qj'}$. The $\Gamma_{qj}$ and $\Gamma_{qj'}$ increase gradually from 0 along the direction of the arrow. The value magnitude is rendered by a gradient color, where yellow areas represent higher values and dark colors represent lower values. The purple line represents the function $\Gamma _{qj+qj'} = |\Delta \omega_{qjj'}|$, as well as the maximum value in the heat map. The inset shows three pairs of Lorentz functions. Here, we assume that each pair of Lorentz functions has the same line width under the three cases with different values. } 
    \end{center}
 \end{figure}

\subsection{Coherences' thermal conductivity}
\par We next move on to unravel the microscopic mechanism of cohereces' thermal conductivity $\kappa^C_{L}$. As shown in Fig. 4(c), a criterion proposed by Michelle et al.~\cite{simoncelli2019unified, simoncelli2022wigner}—the Wigner limit in time (
$ \tau_{Wigner}= 3N_{at}/\omega_{max}$), where $N_{at}$ is the number of atoms and $\omega_{max}$ is the maximum phonon frequency)—can be used to distinguish the particle-like and wave-like nature of phonons. More specifically, phonons with lifetime $\tau_{qj} > \tau_{Wigner}$
  primarily contributes to the populations' thermal conductivity, while phonons with lifetimes $\tau_{qj} < \tau_{Wigner}$ mainly govern the coherences' thermal conductivity~\cite{simoncelli2019unified}. Similarly, the averaged interatomic spacing distance can also be used to distinguish particle-like and wave-like phonons in a frequency-dependent mean-free path plot (see the inset in Fig. 4(c)). As illustrated in Fig. 4(c), most phonons clearly exhibit lifetimes shorter than the Wigner time limit~\cite{simoncelli2022wigner}, suggesting that phonon thermal transport predominantly occurs in a diffusive-like manner.  
   Indeed, the coherent $\kappa^C_{L,z}$ along the $z$-axes dominates thermal transport in crystalline CsCu$_4$Se$_3$, contributing to 74\% of the total $\kappa_{L}$ at 300 K. Although the particle-like $\kappa^P_{L,x}$
  along the $x$-axis is the dominant contributor to the total $\kappa_{L}$, the coherence contribution $\kappa^C_{L,x}$ is substantial, accounting for 43\% at 300 K (see Fig. 4(b)).

 \begin{figure*}[t]
    \begin{center}
    \includegraphics[height=10cm]{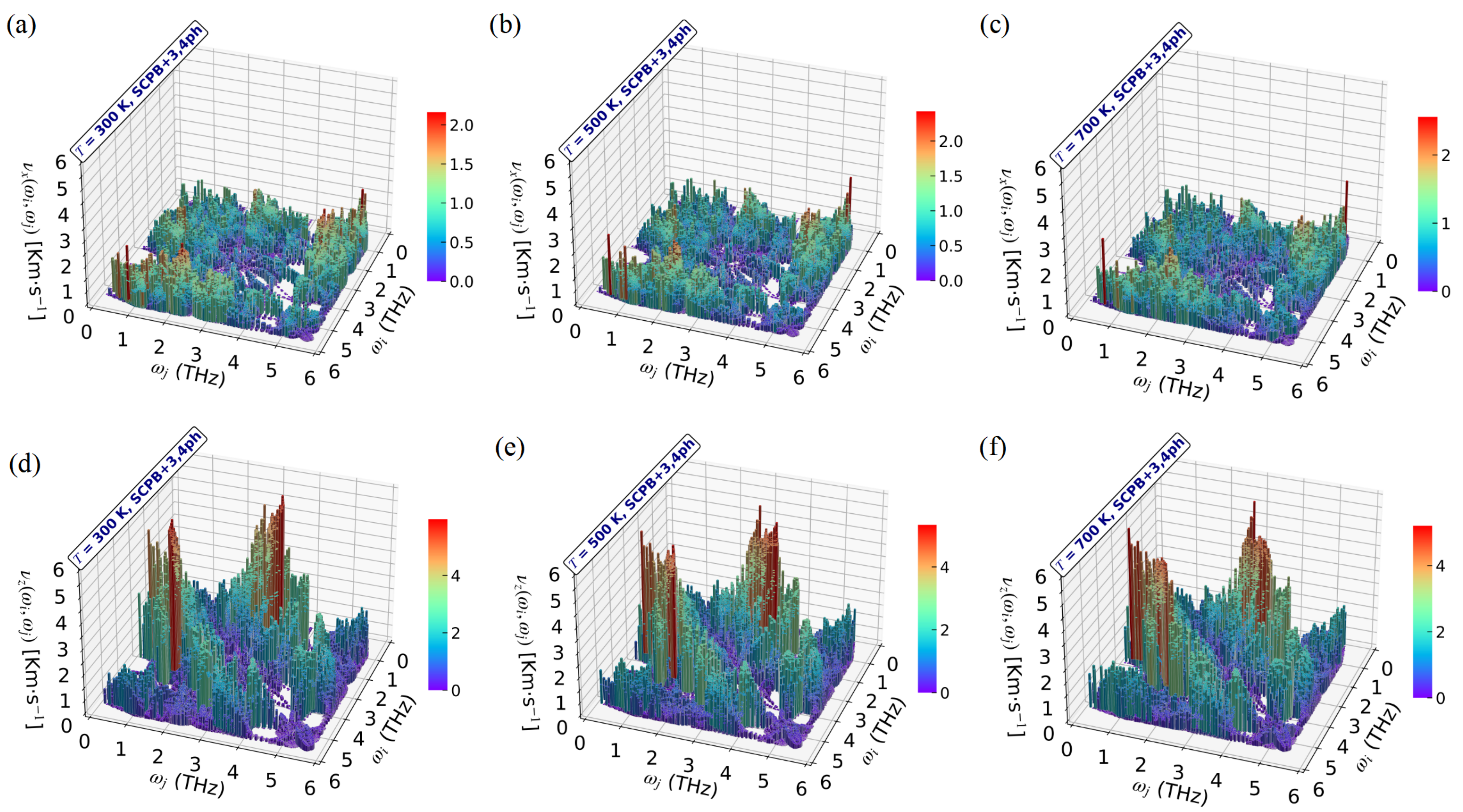}
    \caption{(a) Three-dimensional visualizations of the interband group velocity $\upsilon$($\omega_{qj}$,$\omega_{qj'}$) based on the SCPB+3,4ph model along with the $x$-axis at 300 K. The diagonal data points ($\omega_{qj}$ = $\omega_{qj'}$) indicate phonon degenerate eigenstates. (b) The same as (a), but for 500 K. (c) The same as (a), but for 700 K. (d) The same as (a), but for $z$-direction. (e) The same as (d), but for 500 K.
    (f) The same as (d), but for 700 K.} 
    \end{center}
 \end{figure*}

\par We further conduct an in-depth analysis of the impact of strong quartic anharmonicity on the coherences' thermal conductivity 
$\kappa^C_{L}$. As depicted in Figs. 4(a-b), both $\kappa^C_{L,x}$ and $\kappa^C_{L,z}$ exhibit an upward trend due to additional 4ph scattering, increasing from 0.11 Wm$^{-1}$K$^{-1}$ and 0.46 Wm$^{-1}$K$^{-1}$ to 0.12 Wm$^{-1}$K$^{-1}$ and 0.5 Wm$^{-1}$K$^{-1}$, respectively. Importantly, we observe several notable drop points (1.2 THz, 2 THz, 2.5 THz, 3 THz) induced by 4ph scattering in Fig. 4(c). These drop points correspond to raised peaks at the same frequencies in the spectral/cumulative coherences' wave-like conductivity $\kappa^C_{L,z}(\omega)$ (see Fig. 4(f)). In particular, the coherences' wave-like conductivity increased significantly by 28\% within the frequency range of 0.5–1.2 THz. As mentioned earlier, this region is primarily governed by the Cu-dominated acoustic branch with strong quartic anharmonicity, confirming that additional 4ph scattering positively contributes to $\kappa^C_{L,z}$ (see Fig. 3(c)). 

\par In fact, there is a nonmonotonic relationship between the
coherent contribution and the anharmonic scattering rates, as
shown in Fig. 6 (The code generating this heatmap is provided in SM~\cite{Supplemental}). Mathematically, from Eq. (5), the coherences' thermal conductivity for two specific phonon modes $j$ and $j'$ is proportional to $\Gamma _{qj+qj'}/(|\Delta\omega_{qjj'}|^2+(\Gamma _{qj+qj'})^2)$), where $\Gamma _{qj+qj'}$ is the total linewidth and $\Delta\omega_{qjj'}$ is the frequency difference.
More specifically, if the total linewidths are smaller than the frequency difference between the two eigenstates,i.e., $\Gamma _{qj+qj'} < |\Delta\omega_{qjj'}|$, there is a positive contribution to $\kappa^C_L$ when additional scattering is considered; otherwise, wave-like tunneling between eigenstates will be suppressed [See Fig. 6]. Physically, for two decaying phonon modes to contribute coherently to lattice thermal conductivity, their Lorentzian-shaped spectral functions $b(qj,\omega)$ ($b(qj,\omega) = \Gamma _{qj}/((\omega-\omega_{qj})^2+\Gamma _{qj}^2)$) must overlap~\cite{caldarelli2022many, dangic2024lattice}. The maximum integral value between two Lorentzian functions occurs when the frequency difference between the phonons equals the sum of their linewidths (i.e., the $\Delta\omega_{qjj'}$ in Fig. 6). Additionally, the amplitude of the Lorentzian-shaped spectral function of phonons, which represents the likelihood of specific phonon states, can be interpreted as the efficiency of coherent thermal transport between the two modes. When phonon linewidths are extremely large, the amplitude of the Lorentzian function decreases, indicating a lower probability for specific phonon states (See the illustration of Fig. 6). To intuitively quantify this trend, we calculated integral values (I) of two Lorentzian functions with the
same linewidths and fixed frequency difference $\Delta\omega_{qjj'}$, including three cases with linewidth values follow by $\Gamma_{case1} < \Gamma_{case2} = \Delta\omega_{qjj'} < \Gamma_{case3}$. The resulting integral values are labeled as $\mathrm{I_1}$, $\mathrm{I_2}$, and $\mathrm{I_3}$, and their relative relationship also follow by $\mathrm{I_2} > \mathrm{I_1}$ and $ \mathrm{I_2} > \mathrm{I_3}$. As a result, large phonon scattering rates (or linewidths) do not always lead to a large coherence contribution; the coherence contribution also depends on the amplitude of the spectral function.

\par Interestingly, we observe that the $z$-axis coherences' thermal conductivity exhibits non-monotonic behavior with increasing temperature, as shown in Fig. 4(b). For crystalline CsCu$_4$Se$_3$, the $\kappa^C_L$ experiences a significant increase with the temperature initially rising, which corresponds to the condition of total linewidth growth below $|\Delta\omega_{qjj'}|$ is well satisfied. 
Then, the $\kappa^C_L$ reaches its maximum at 500 K, corresponding to the condition where most pairs of eigenstates follow by $|\Delta\omega_{qjj'}| \simeq \Gamma _{qj+qj'}$. As the temperature increases, the scattering of phonons becomes significant and their corresponding Lorentzian amplitude reduces, which in turn suppresses the wave-like tunneling behavior between pairs of vibrational eigenstate and leads to a decrease in $\kappa^C_L$ (see Fig. 4(b)). Notably, it is important to underscore that the harmonic characteristics of phonon undergo certain modifications due to renormalization, which directly impacts the generalized interband group velocity between two eigenstates. As a result, the non-monotonic temperature dependence of coherences' thermal conductivity results from the combined effects of anharmonic scattering rates and anharmonic phonon renormalization.

\par Then, we delve into the anisotropic behavior of $\kappa^C_{L}$ by
analyzing the spectral and cumulative coherences’ thermal conductivity, as shown in Fig. 4(f). In Fig. 4(f), the contribution of $\kappa^C_{L,z}$ comes from a wide range of the spectral distribution, while the contribution of $\kappa^C_{L,x}$ is primarily concentrated in the low-frequency range (1-2.5 THz). To intuitively understand the anisotropic behavior of coherence contribution, we plot the three-dimensional visualizations of the coherences' thermal conductivity $\kappa^C_L$($\omega_{qj}$,$\omega_{qj'}$), as demonstrated in Figs. 5(a-f). The results illustrate that the contribution of $\kappa^C_{L,x}$($\omega_{qj}$,$\omega_{qj'}$) mainly comes from the coupling between quasi-degenerate states ($\omega_{qj} \simeq \omega_{qj'}$), similar to the case for harmonic glasses ($\Gamma_q \to 0$)~\cite{simoncelli2019unified}. Conversely, the non-degenerate phonons in the 1-1.5 THz frequency range, driven by strong anharmonicity, mainly contribute to $\kappa^C_{L,z}$($\omega_{qj}$,$\omega_{qj'}$). 
Besides, we consider generalized interband group velocity $\upsilon$($\omega_{qj}$,$\omega_{qj'}$) to further understand anisotropy, as depicted in Figs. 7(a-f). Along the $x$-direction, non-degenerate states with large frequency discrepancy, exhibit higher interband group velocity $\upsilon_x$($\omega_{qj}$,$\omega_{qj'}$). However, the large frequency discrepancy, in turn, hinders the coherence effect since it needs larger $\Gamma_{qj+qj'}$ or an external perturbation to drive the interband tunneling~\cite{caldarelli2022many}.
Therefore, the majority of $\kappa^C_{L,x}$ primarily arises from quasi-degenerate states ($\omega_{qj} \simeq \omega_{qj'}$) with lower interband velocities, rather than from non-degenerate states characterized by large frequency discrepancy (see Figs. 5(a-c)). In comparison, the distribution of $\upsilon_z$($\omega_{qj}$,$\omega_{qj'}$) aligns closely with that of $\kappa^C_{L,z}$($\omega_{qj}$,$\omega_{qj'}$) (see Figs. 5(d-f)). Many non-degenerate states with moderate frequency differences possess higher interband velocities, which leads to higher $\kappa^C_{L,z}$. As a result, although the particle-like thermal propagation would be hindered by the hierarchical configuration along the $z$-direction, an effective cross-layer thermal transport can still be achieved via the wave-like tunnelings between the eigenstates.

\section{Conclusions}
\par In summary, we have systematically investigated the lattice dynamics associated with higher-order anharmonic self-energies and the thermal transport properties of crystalline CsCu$_4$Se$_3$ using a unified thermal transport theory. Using the harmonic approximation approach, we observe unstable modes in crystalline CsCu$_4$Se$_3$, indicating dynamical instability and strong anharmonicity. These unstable soft modes are induced by the tilting motion of the Cu$_4$Se units, and their double-well potential energy surfaces can be accurately captured by incorporating higher-order force constants. To obtain the finite-temperature phonon dispersions, we used a self-consistent theory that incorporates bubble and loop diagrams to non-perturbatively account for the frequency shifts. In stable phonon dispersions, the Cu atoms dominate the low-frequency flat phonon branches, while Cs atoms exhibit highly localized behavior. However, we find that both Cu- and Cs-dominated phonon modes exhibit strong anharmonic scatterings, which play a major role in suppressing lattice thermal conductivity in crystalline CsCu$_4$Se$_3$.

\par Using a unified theory of thermal transport that incorporates both diagonal and off-diagonal terms of heat flux operators, we predict ultra-low thermal conductivity values of 0.39 Wm$^{-1}$K$^{-1}$ and 0.67 Wm$^{-1}$K$^{-1}$ along the $x$ and $z$ axes, respectively. We find that the out-of-plane thermal conductivity is higher than the in-plane thermal conductivity, which is in sharp contrast to what is typically observed in other layered materials. This observation can be explained by the large coherence contributions along the $z$-axis, indicating that heat transport can be effectively facilitated across layers through wave-like tunneling driven by the coupling of distinct vibrational eigenstates. More specifically, the large coherent wave-like conductivity along the $z$-axis contributes to 74\% of the total $\kappa_L$ at 300 K.
Meanwhile, the dominance of coherence contributions results in an anomalous, glassy-like thermal transport behavior over a wide temperature range (100-700 K) along the 
$z$-axis. Finally, we demonstrated that the non-monotonic temperature dependence of coherences' conductivity arises from combined effects between higher-order anharmonic scattering and anharmonic
phonon renormalization.
Our study not only clarifies the crucial role of coherences' thermal conductivity in cross-layer thermal transport but also provides an in-depth analysis of the non-monotonic temperature dependence of its glass-like thermal behavior.

\begin{acknowledgments}
J.Yue and Y.Liu acknowledge the National Natural Science Foundation of China through grants No. 52072188 and No. 12204254, and the Program for Science and Technology Innovation Team in Zhejiang through grant No. 2021R01004, as well as the Institute of High-pressure Physics of Ningbo University for its computational resources. We also acknowledge Ðorđe Dangić from the University of the Basque Country for the valuable and insightful discussions.
\end{acknowledgments}

\bibliography{ref}

\end{document}